\documentclass{aa}
\usepackage{psfig}
\def\gtrsim{\mathrel{\hbox{\rlap{\hbox{\lower4pt\hbox{$\sim$}}}\hbox{$>$}}}}

\def\lesssim{\mathrel{\hbox{\rlap{\hbox{\lower4pt\hbox{$\sim$}}}\hbox{$<$}}}}
\let\lsim=\lesssim

\begin{document}

\title{GRB Redshift determination in the X--ray band}
\authorrunning{Campana et al.}

\author{S. Campana\inst{1}, G. Ghisellini\inst{1}, D. Lazzati\inst{1,2}, 
F. Haardt\inst{3}, \and S. Covino\inst{1}}

\institute{Osservatorio Astronomico di Brera, Via Bianchi 46, I--23807
Merate (Lc), Italy
\and
Dipartimento di Fisica, Universit\`a degli Studi di Milano,
Via Celoria 16, I--20133 Milano, Italy
\and
Dipartimento di Fisica dell'Universit\`a di Como, via Lucini 3, 
I--22100 Como, Italy}

\maketitle

\begin{abstract}
If $\gamma-$ray bursts originate in dense stellar forming regions, the
interstellar material can imprint detectable absorption features on the observed
X--ray spectrum.  Such features can be detected by existing and planned X--ray
satellites, as long as the X--ray afterglow is observed after a
few minutes from the burst. Detection of these X--ray features will
make possible the determination of the redshift of $\gamma-$ray bursts
even when their optical afterglows are severely dimmed by extinction.
For further detail see Ghisellini et al. (1999).
\end{abstract}


\section{The $N_{\rm H}$--redshift plane}

If $\gamma$-ray bursts (GRBs) originate in a dense environment their X--ray
afterglow spectra are modified by absorption features and the imprinted edges can
be used to determine their redshifts. Given the time decay law observed in the
GRB X--ray afterglows, the necessary S/N ratio to reveal an absorption feature
can be achieved only if the X--ray observation starts immediately after the burst
itself, or if the collecting effective area of the detector is much larger than
100 cm$^2$.

We have simulated the observed spectrum by using the response matrices of some
future planned missions, such as JET-X ($A\sim 200$ cm$^2$ at 1.5 keV and $A\sim
40$ cm$^2$ at 8.1 keV for the two telescopes; Citterio et al. 1996), AXAF with
Back Illuminated (BI) CCDs ($A\sim 700$ cm$^2$ at 1.5 keV and $A\sim 40$ cm$^2$
at 8.1 keV; Kellogg et al. 1997) and XMM with the EPIC detectors ($A\sim 3600$
cm$^2$ at 1.5 keV and $A\sim 1500$ cm$^2$ at 8.1 keV for three telescopes;
Gondoin et al. 1996) and assuming:  $i)$ $F(100\,{\rm s})=10^{-8}$ erg cm$^{-2}$
s$^{-1}$ between 2 and 10 keV at the beginning of the observation;  $ii)$ a power
law time decay of the flux $\propto t^{-1}$;  $iii)$ an intrinsic (unabsorbed)
power law spectrum of photon index $\Gamma=1$ constant in time.  All the
simulations reported here refer to observations of 10 ks.

We simulated two different cases: a GRB afterglow at $z=0.25$ and intrinsic
$N_{\rm H}=3\times 10^{21}$ cm$^{-2}$ and $z=4$ and $N_{\rm H}=10^{24}$
cm$^{-2}$, which are relevant for the oxygen and iron edge, respectively.  A
galactic column density of $3\times 10^{20}$ cm$^{-2}$ has also been included
(for an overview of the $N_H$ values with BeppoSAX see Owens et al.
1998).
In the case of the oxygen edge (at 0.52 keV) the satellite energy band is
extremely important in order to recover the correct GRB redshift. We keep
fixed the edge energies, even if in the case of a warm absorber fit should be
worse.

In the case of JET-X, the minimum energy of 0.3 keV limits the maximum detectable
redshift to $\sim 0.7$. The influence of the galactic absorption plays also a
crucial role, such that only for low values ($\lsim\, 5\times 10^{20}$ cm$^{-2}$)
we are able to disentangle the intrinsic and the galactic absorption.

In Fig. 1 (left side) we report the contour plots in the $N_{\rm H}-z$ plane of
the simulated models as observed with different X--ray satellites.  The three
contours refer to 1, 2 and $3\,\sigma$ confidence levels.
In Fig. 1a is shown the case of the JET-X telescope.  It can be noted that the
input redshift and column density are not recovered satisfactorily.  In
particular, the presence of different absorption features (O, Ne, Mg, Si) results
in the elongated contour in the $N_{\rm H}-z$ plane.  In the case of AXAF (Fig.
1b), the recovery of the GRB redshift is eased by the higher throughput at low
energies guaranteed by the BI CCDs.
The large effective area of XMM poses no problem for the identification of the
redshift (Fig. 1c).

In the case of the Fe edge there are less problems due to the fact that beyond
iron there are not prominent K edges.  This is testified by Fig. 1 (right side),
in which for all the considered instrument the redshift and the column density
are recovered with a high degree of confidence. Note however that at these
large redshift, the iron abundance may be lower than the solar value.

\begin{figure*}
\centerline{\psfig{figure=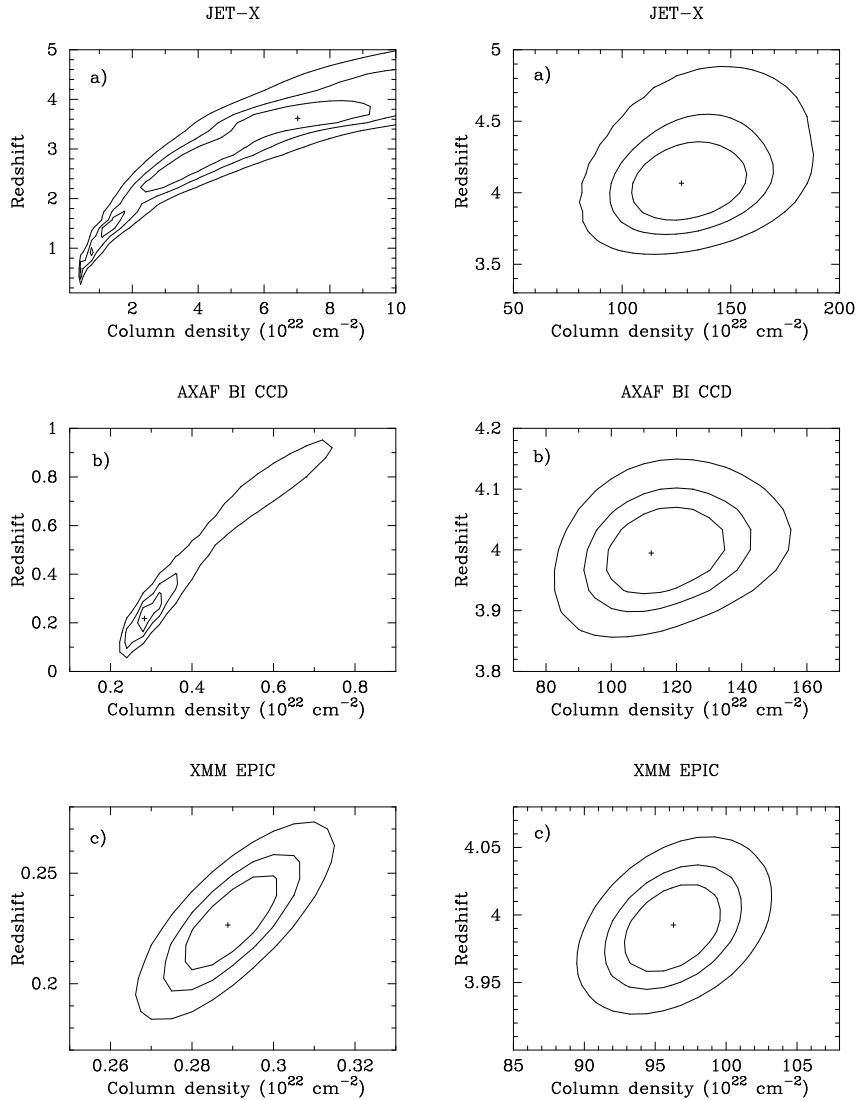,height=16. truecm}}
\caption[h]{Column density -- redshift contour plots for different X--ray
instruments. LEFT SIDE: The input model has $z=0.25$ and $N_{\rm H}=3\times
10^{21}$ cm$^{-2}$.  The upper panel (a) shows the case of JET-X.  The middle
panel (b) presents the case of AXAF with BI CCDs and the lower panel (c) the case
of XMM.  RIGHT SIDE: The input model has $z=4$ and $N_{\rm H}=10^{24}$
cm$^{-2}$.}
\end{figure*}

\section{Discussion}

Oxygen and iron edges are the most prominent absorption features in the spectra
of X--ray sources. This individuates two almost distinct accessible part of the
redshift--column density plane for GRB:
one characterized by a moderate $N_{\rm H}\sim 10^{21}$--$10^{22}$ cm$^{-2}$ and
$z\sim 0.1$--$0.5$ and the other one by a column larger than $10^{23}$ cm$^{-2}$.
Note that for a standard dust to gas ratio, a column of $N_{\rm H} =10^{22}$
cm$^{-2}$ corresponds to an optical extinction $A_V\sim 6$ mag, precluding the
possibility to detect the optical afterglow.
If, furthermore, it were not possible to perform spectroscopic optical
observation of the host galaxy (either because a precise position is lacking, or
because it is too faint), then the X--rays could be the only mean to determine
the redshift, especially for strong bursts located at $z > 2-3$, for which the
iron edge lies in the most sensitive energy range of X--ray detectors.

\end{document}